\documentstyle[12pt,epsf]{article}
\def\bm#1{\mbox{\boldmath $#1$}}
\def\it#1{\mbox{$#1$}}

\def\beq{\begin{equation}}
\def\eeq{\end{equation}}

\def\t2{\mbox{  }}
\def\rst1{\mbox{ }}

\def\frc#1,#2{{{#1}\over{#2}}}
\begin{document}
\setcounter{equation}{0}
\renewcommand{\theequation}{\arabic{equation}}
\renewcommand{\thesection}{\Roman{section}.}
\renewcommand{\thesubsection}{\Alph{subsection}.}
\renewcommand{\thesubsubsection}{\arabic{subsubsection}.}

\title{Holonomic constraints : an analytical result.}
\author{\normalsize Martial MAZARS\footnote{Electronic mail: Martial.Mazars@th.u-psud.fr} \\
\small Laboratoire de Physique Th\'eorique (UMR 8627),\\
\small Universit\'e de Paris Sud XI, B\^atiment 210, 91405 Orsay Cedex,
FRANCE}
\maketitle

\hfill\small\hspace{3.0in} L.P.T.-Orsay   07-01      \\

\begin{center}{\bf Abstract}\end{center}
Systems subjected to holonomic constraints follow quite complicated dynamics that could not be described easily with Hamiltonian or Lagrangian dynamics. The influence of holonomic constraints in equations of motions is taken into account by using Lagrange multipliers. Finding the value of the Lagrange multipliers allows to compute the forces induced by the constraints and therefore, to integrate the equations of motions of the system. Computing analytically the Lagrange multipliers for a constrained system may be a difficult task that is depending on the complexity of systems. For complex systems, it is  most of the time impossible to achieve. In computer simulations, some algorithms using iterative procedures estimate numerically Lagrange multipliers or constraint forces by correcting the unconstrained trajectory. In this work, we provide an analytical computation of the Lagrange multipliers for a set of linear holonomic constraints with an arbitrary number of bonds of constant length. In the appendix of the paper, one would find explicit formulas for Lagrange multipliers for systems having 1, 2, 3, 4 and 5 bonds of constant length, linearly connected.    
\newpage

\section{Introduction.}

The characteristic time scales associated with intramolecular motions are around 10 to 100 times shorter than the charecteristic time scales associated with translational and rotational degrees of freedom of the molecule. In Molecular Dynamic computations a convenient way to handle the multiple-time-scale in equation of motions is to treat covalent bonds between atoms as rigid, this procedure introduces constraints on equations of motions ; the use of the Lagrange multipliers method allows to integrate the equations of motion \cite{r1}.\\
From a technical point of view, the Lagrange multipliers method is used in SHAKE and RATTLE algorithms : the constrained trajectories of the systems are computed from unconstrained trajectories, by using an iterative algorithm that allows to estimate numerically the Lagrange multipliers \cite{r2,r3,r4}. These algorithms are very efficient and quite convenient. Nevertheless, a closed analytical computation of the Lagrange multipliers is interesting on its own on a theoretical ground and can also be used to improve the iterative estimations done in SHAKE and RATTLE algorithms.\\ 
The equations of motions of the constrained dynamics are derived as follow. For a set of holonomic constraints, noted as $\sigma_{\alpha}\equiv 0$, one defines a constrained Lagrangian $L_{c}[q,\dot{q}]$ as
\beq
L_{c}[q,\dot{q}]=L[q,\dot{q}]-\sum_{\alpha}\lambda_{\alpha}\sigma_{\alpha}(q,\dot{q})
\eeq
where $L[q,\dot{q}]$ is the Lagrangian of the unconstrained systems and $\lambda_{\alpha}$ the set of Lagrange multipliers. The equations of motions are given by 
\beq
\frac{\partial}{\partial t}\frac{\partial L'}{\partial \dot{q}}=\frac{\partial L'}{\partial q}
\eeq
which may also be written as
\beq
m_{i}\ddot{q}_{i}=F_{i}+\sum_{\alpha}\lambda_{\alpha}\frac{\partial \sigma_{\alpha}}{\partial q_i}=F_{i}+\sum_{\alpha}G_{i;\alpha}
\eeq
where the constraint forces are defined by $G_{i;\alpha}$. Since holonomic constraints are conserved quantities, by requiring that the second time derivatives of all $\sigma_{\alpha}$ vanish, we obtain a set of closed equations that allows to compute the Lagrange multipliers. We have
\beq
\begin{array}{ll} 
\displaystyle \frac{\partial\dot{\sigma_{\alpha}}}{\partial t}&= \displaystyle\sum_{i}\frac{1}{m_i}\mbox{\large (}F_i+\sum_{\beta}G_{i;\beta} \mbox{\large )}\nabla_{i}\sigma_{\alpha}+\sum_{i,j}\dot{q_i}\dot{q_j}\nabla_{i,j}\sigma_{\alpha} \\
&\\
&= \displaystyle K_\alpha+C_\alpha+\sum_{\beta}Z_{\alpha \beta}\lambda_{\beta}=0
\end{array}
\eeq
where we have set 
\beq
\left\{
\begin{array}{ll}
K_\alpha &= \displaystyle \sum_{i}\frac{1}{m_i}F_i\cdot\nabla_{i}\sigma_{\alpha}\\
&\\
C_\alpha &= \displaystyle \sum_{i,j}\dot{q_i}\dot{q_j}\nabla_{i,j}\sigma_{\alpha} \\
&\\
Z_{\alpha \beta}&= \displaystyle \sum_{i}\frac{1}{m_i}\nabla_{i}\sigma_{\alpha}\cdot\nabla_{i}\sigma_{\beta}
\end{array}\right .
\eeq
Then, Lagrange multipliers are given by finding the inverse of the Z-matrix, as
\beq
\lambda_\beta=-\sum_{\alpha}(Z^{-1})_{\beta\alpha}(K_\alpha+C_\alpha)
\eeq
Obviously, the structure of the Z-matrix is closely related to the geometrical features of molecules, its computation can be quite complicated and is depending on the system.\\ 
In this work, by computing the inverse of the Z-matrix, we provide an analytical result for holonomic constraints involved in the dynamic of linear molecules (polymers, etc.) or having a linear sequence in its structure (branched polymers, etc). For such systems, the holonomic constraints can be read as
\beq
\displaystyle\sigma_{\alpha}(\bm{r}_{\alpha},\bm{r}_{\alpha-1})\equiv (\bm{r}_{\alpha}-\bm{r}_{\alpha-1})^{2}-a^{2}=0
\eeq
where $\bm{r}_{\alpha}$ are the coordinates of the atom numbered $\alpha$ in the linear sequence of the molecule and $a$ the length of the bond between atoms $\alpha$ and $\alpha-1$. $N$ is the number of holonomic constraints, these constraints involve $N$ bonds and $N+1$ atoms. \\
In the next section, we provide a closed analytical computation of the Z$^{-1}$-matrix for the linear holonomic constraints given by Eq.(7).

\section{Analytical computation of Lagrange multipliers for linear holonomic constraints.}

We set $\bm{r}_{\alpha}-\bm{r}_{\alpha-1}=a\hat{u}_{\alpha}$, where $\hat{u}_{\alpha}$ is the unitary bond vector linking the atom numbered $\alpha-1$ to atom $\alpha$ ; the linear holonomic constraints Eq.(7) gives
\beq
\nabla_{i}\sigma_{\alpha}=2a(\delta_{i,\alpha}-\delta_{i,\alpha-1})\hat{u}_{\alpha}
\eeq
where we use the Kronecker symbol $\delta_{n,p}$. The force acting on atom $i$ due to all holonomic constraints is $G_i=\sum_{\alpha}G_{i;\alpha}$ and we found easily that 
\begin{center}
$\displaystyle \sum_i G_i=0$
\end{center}
This may be easily verified for linear holonomic constraints.\\ 
Thus, for a set of linear holonomic constraints, the Z-matrix is given by 
\beq
\begin{array}{ll}
\displaystyle Z_{\alpha\beta}&= \displaystyle\frac{2a^2}{m}\sum_{i}(\delta_{i,\alpha}-\delta_{i,\alpha-1})(\delta_{i,\beta}-\delta_{i,\beta-1})\hat{u}_{\alpha}\hat{u}_{\beta}\\
&\\
&= \displaystyle \frac{4a^2}{m}\mbox{\large (}-\frac{1}{2}(\hat{u}_{\alpha-1}\cdot\hat{u}_{\alpha})\delta_{\alpha,\beta+1}+\delta_{\alpha,\beta}-\frac{1}{2}(\hat{u}_{\beta-1}\cdot\hat{u}_{\beta})\delta_{\alpha,\beta-1}\mbox{\large )}
\end{array}
\eeq
In the following, we set
\beq
\gamma_{\alpha}=\frac{1}{2}\hat{u}_{\alpha-1}\cdot\hat{u}_{\alpha}\t2\t2\t2\mbox{and}\t2\t2\t2\tilde{Z}_{\alpha\beta}=-\gamma_{\alpha}\delta_{\alpha,\beta+1}+\delta_{\alpha,\beta}-\gamma_{\beta}\delta_{\alpha,\beta-1}
\eeq
In appendix, computations for $N=1,2,3,4$ and 5 are given explicitly. With the linear holonomic constraints the Z-matrix is a banded symmetric matrix. To compute the inverse matrix we take into account the properties of $\tilde{Z}$. The inverse matrix is built by a sequence of multiplications on the right and the left in such a way that the identity matrix appears progressively on the diagonal. More precisely, according to Eq.(10), if we perform the transform 
\beq
\tilde{Z}\longrightarrow B_{1}^{t}A_{1}^{t}\tilde{Z}A_{1}B_{1}=\tilde{Z}^{(1)}
\eeq
with matrix $A_1$ and $B_1$ defined by 
\beq
\displaystyle (A_1)_{ij}=\delta_{i,j}+\gamma_{2}\delta_{i,1}\delta_{2,j}\t2\t2\t2\mbox{and}\t2\t2\t2(B_1)_{ij}=\delta_{i,j}+\mbox{\Large (}\frac{1}{\sqrt{1-\gamma_{2}^{2}}}-1\mbox{\Large )}\delta_{i,2}\delta_{2,j}
\eeq
then, the $2\times 2$ identity matrix appears in $\tilde{Z}^{(1)}$ for $1\leq i\leq2$ and $1\leq j\leq2$. The matrix multiplications by $A_{1}$ and $A_{1}^{t}$ result in vanishing the non-diagonal coefficient (i.e. $\tilde{Z}_{12}=\tilde{Z}_{21}=-\gamma_2$), while matrix multiplications by $B_{1}$ and $B_{1}^{t}$ scale to 1 the diagonal coefficient  $(A_{1}^{t}\tilde{Z}A_{1})_{22}$. The matrix $\tilde{Z}^{(1)}$ is similar to the original matrix $\tilde{Z}$, with $\gamma_{2}$ and $\gamma_{3}$ transformed as
\beq
\begin{array}{ll}
\displaystyle \gamma_{2}\longrightarrow& \gamma'_{2}=0\\
&\\
\displaystyle \gamma_{3}\longrightarrow&\displaystyle \gamma'_{3}=\frac{\gamma_{3}}{\sqrt{1-\gamma_{2}^{2}}}
\end{array}
\eeq
because of Eq.(11).\\
We are then in a position to build the identity matrix from $\tilde{Z}$, by transformations similar to Eq.(11). The matrix $\tilde{Z}^{(n)}$ is computed from $\tilde{Z}^{(n-1)}$ by making
\beq
\tilde{Z}^{(n)}=B_{n}^{t}A_{n}^{t}\tilde{Z}^{(n-1)}A_{n}B_{n}
\eeq
With a convenient choice of the matrix $A_{n}$ and $B_{n}$, we will finally obtain $\tilde{Z}^{(N-1)}=I_N$ and therefore, we will have built a matrix $C$ such as
\beq
C^{t}\tilde{Z}C=I_N\t2\t2\t2\mbox{with}\t2\t2\t2C=\prod_{n=1}^{N-1}(A_nB_n)
\eeq
and thus
\beq
\tilde{Z}^{-1}=CC^{t}
\eeq
To this end, taking into account the structure of Eq.(13) for $\gamma'_{3}$, we define the sequence $\Omega_n$ by
\beq
\Omega_n^2=1-\frac{\gamma_n^2}{\Omega_{n-1}^{2}}
\eeq 
with $\gamma_1=0$, $\Omega_1=1$ and, for convenient reasons, $\gamma_{N+1}=0$, that gives  $\Omega_{N+1}=1$. With these notations the matrix $A_n$ and $B_n$ are given by  
\beq
\begin{array}{ll}
\displaystyle (A_n)_{ij}&=\displaystyle\delta_{i,j}+\frac{\gamma_{n+1}}{\Omega_n}\delta_{i,n}\delta_{(n+1),j}\\
&\\
\displaystyle (B_n)_{ij}&=\displaystyle\delta_{i,j}+\mbox{\Large (}\frac{1}{\Omega_{n+1}}-1\mbox{\Large )}\delta_{i,(n+1)}\delta_{(n+1),j}
\end{array}
\eeq
From the definition of matrix $A_n$ and $B_n$ and with Eq.(16), we may easily compute determinants as
\begin{center}
$\displaystyle \mbox{Det}(A_n)\displaystyle=1\t2\t2\t2\t2 ;\t2\t2\t2\t2\mbox{Det}(B_n)\displaystyle=\frac{1}{\Omega_{n+1}}$
\end{center}
and
\beq
\mbox{Det}(\tilde{Z}^{-1})\displaystyle=\prod_{n=1}^{N-1}\frac{1}{\Omega_{n+1}^2}=\frac{1}{\mbox{Det}(\tilde{Z})}
\eeq
$\mbox{Det}(\tilde{Z})$ is closely related to the metric determinant that is used in computations of ensemble averages \cite{r5,r6,r6a}.\\
Eqs.(16) and (17) allow to compute the $\tilde{Z}^{-1}$ matrix coefficients, the matrix is symmetric $\tilde{Z}^{-1})_{ij} = \tilde{Z}^{-1})_{ji}$, and after some algebra we found
\beq
\left\{
\begin{array}{lll}
\mbox{For $i=j$}:\t2\t2\t2&\displaystyle \tilde{Z}^{-1}\mbox{\Large )}_{ii} &\displaystyle=\frac{1}{\Omega_i^2}\mbox{\Large [}1+\sum_{p=i+1}^{N}\prod_{m=i+1}^{p}\mbox{\large (}\frac{1}{\Omega_m^2}-1\mbox{\large )}\mbox{\Large ]}\\
&\\
\mbox{For $i<j$}:\t2\t2\t2&\displaystyle \tilde{Z}^{-1}\mbox{\Large )}_{ij} &\displaystyle=\frac{1}{\Omega_i^2}\mbox{\Large [}\prod_{n=i+1}^{j}\frac{\gamma_n}{\Omega_n^2}\mbox{\Large ]}\mbox{\Large [}1+\sum_{p=j+1}^{N}\prod_{m=j+1}^{p}\mbox{\large (}\frac{1}{\Omega_m^2}-1\mbox{\large )}\mbox{\Large ]}
\end{array}
\right .
\eeq
With Eqs.(10) and (20), we may verify that $\tilde{Z}^{-1}\tilde{Z}=I_N$. Eqs.(19) and (20) are the main results of this paper.\\
According Eq.(6), we have to compute $K_\alpha$ and $C_\alpha$ to achieve the analytical computation of the Lagrange multipliers. With Eqs.(5) and (8), we found
\beq
K_\alpha=\frac{2a}{m}(\bm{F}_\alpha-\bm{F}_{\alpha-1})\cdot\hat{\bm{u}}_{\alpha}
\eeq
and
\beq
C_\alpha=\frac{2}{m^2}(\bm{p}_\alpha-\bm{p}_{\alpha-1})^2
\eeq
with the momentum of atoms, $\bm{p}_\alpha=m\bm{v}_\alpha=m\dot{\bm{r}}_\alpha$ and notations defined previously. Lagrange multipliers are then given by applying Eq.(6).\\
So far, we have considered that all atoms, involved in the set of holonomic constraints, have the same mass $m$ and the length of each bond is equal to $a$. We consider now that the sequence of atoms mass is $\{m_i\}_{i\in [0,N]}$ and the length of bonds are $\{a_i\}_{i\in [1,N]}$. Then, the momentum of atoms is now $\bm{p}_\alpha=m_{\alpha}\bm{v}_\alpha=m_{\alpha}\dot{\bm{r}}_\alpha$ and Eq.(8) becomes
\beq
\nabla_{i}\sigma_{\alpha}=2a_{\alpha}(\delta_{i,\alpha}-\delta_{i,\alpha-1})\hat{u}_{\alpha}
\eeq
Therefore, the Z-matrix is now given by
\beq
\left\{
\begin{array}{ll}
\displaystyle Z_{\alpha\beta} &= \displaystyle -\nu_\alpha(\hat{u}_{\alpha-1}\cdot\hat{u}_{\alpha})\delta_{\alpha,\beta+1}+\omega_\alpha\delta_{\alpha,\beta}-\nu_\beta(\hat{u}_{\beta-1}\cdot\hat{u}_{\beta})\delta_{\alpha,\beta-1}\\
&\\
\displaystyle\nu_\alpha &= \displaystyle -\frac{2a_{\alpha-1}a_\alpha}{m_{\alpha-1}}\\
&\\
\displaystyle\omega_\alpha &= \displaystyle 2a_\alpha^2\mbox{\Large (}\frac{1}{m_{\alpha-1}}+\frac{1}{m_{\alpha}}\mbox{\Large )}
\end{array}
\right .
\eeq
Eq.(24) is similar to Eq.(9). We may transform Eq.(24) to Eq.(10) with the help of a diagonal matrix $D$, according to
\beq
\tilde{Z}=DZD
\eeq
with 
\beq
D_{ij}=\frac{1}{\sqrt{\omega_i}}\delta_{ij}
\eeq
Then, $\tilde{Z}$ is again given by Eq.(10), but with
\beq
\gamma_{\alpha}=\sqrt{\frac{m_\alpha m_{\alpha-2}}{(m_{\alpha-1}+m_{\alpha})(m_{\alpha-1}+m_{\alpha-2})}}(\hat{u}_{\alpha-1}\cdot\hat{u}_{\alpha})
\eeq
where inertia parameters of bonds appear explicitly (the canonical partition function of freely jointed chains is depending on inertia parameters of the bonds as defined in ref.\cite{r7}). One may note also that if $m_\alpha=m$, Eq.(27) corresponds to Eq.(10).\\
The inverse of the Z-matrix is now given by
\beq
Z^{-1}=D\tilde{Z}^{-1}D=D(CC^{t})D
\eeq
with Eq.(27), (17) and (20) ; and $K_\alpha$ and $C_\alpha$ are given by
\beq
K_\alpha= 2a_\alpha\mbox{\Large (}\frac{\bm{F}_\alpha}{m_\alpha}-\frac{\bm{F}_{\alpha-1}}{m_{\alpha-1}}\mbox{\Large )}\cdot\hat{\bm{u}}_{\alpha}
\eeq
and
\beq
C_\alpha=2\mbox{\Large (}\frac{\bm{p}_\alpha}{m_\alpha}-\frac{\bm{p}_{\alpha-1}}{m_{\alpha-1}}\mbox{\Large )}^2
\eeq

\section{Discussion.}

From an analytical point of view, an interesting result is given by the computation of the determinant of  $\tilde{Z}$ (Eq.(19)). For holonomic constraints, this determinant is known as the metric determinant ; and for non-hamiltonian dynamical systems \cite{r5,r6,r8} it can be related to the Jacobian determinant, some ensemble averages can be formulated with the help of this determinant \cite{r5,r6,r9}. According to computations done in ref.\cite{r10}, we obtain 
\beq
\begin{array}{ll}
\displaystyle2^{N-1}\int_0^{\frac{1}{2}}\prod_{n=2}^{N}d\gamma_n\sqrt{\mbox{Det}(\tilde{Z}^{-1})}&\displaystyle=2^{N-1}\int_0^{\frac{1}{2}}\prod_{n=2}^{N}d\gamma_n\frac{1}{\sqrt{\mbox{Det}(\tilde{Z}})}\\
&\\
&\displaystyle = \mbox{}^{[N-1]}\mbox{He}^{[N]}(\frac{1}{2} ; \frac{3}{2} ; \frac{1}{4}, ... ,\frac{1}{4})
\end{array}
\eeq
where $\mbox{}^{[N-1]}\mbox{He}^{[N]}$ is the multiple hypergeometric function defined in ref.\cite{r7}, related to the canonical partition function of freely jointed chains. In particular, for $N=2$, Eq.(31) is equivalent to the elementary relation
\beq
2 \int_0^{\frac{1}{2}}\frac{d\gamma_2}{\sqrt{1-\gamma_2^2}}=\mbox{}_2\mbox{F}_1(\frac{1}{2}, \frac{1}{2} ; \frac{3}{2} ; \frac{1}{4}) =2 \arcsin (\frac{1}{2})
\eeq
In ref.\cite{r7}, we have shown that the multiple hypergeometric function $\mbox{}^{[N-1]}\mbox{He}^{[N]}$ is defined for any mass sequence, Eq.(31) may therefore be extended to any mass sequence, and the multiple hypergeometric function $\mbox{}^{[N-1]}\mbox{He}^{[N]}$ must be evaluated at a point defined by the inertia parameters of the coupling between bonds \cite{r7}. The inertia parameter of the coupling between bonds $\alpha$ and $\alpha-1$ is given by
\beq
x_{\alpha}=\frac{m_\alpha m_{\alpha-2}}{(m_{\alpha-1}+m_{\alpha})(m_{\alpha-1}+m_{\alpha-2})}
\eeq
this parameter appears explicitly in Eq.(27), so in the integral too (one may note also that if $m_\alpha=m$, for all $\alpha$, then we have $x_{\alpha}=1/4$). For any mass sequence, the relation equivalent to Eq.(31) is
\beq
\displaystyle\int_0^{\sqrt{x_2}}d\gamma_2\mbox{ ... }\int_0^{\sqrt{x_N}}d\gamma_N\sqrt{\mbox{Det}(\tilde{Z}^{-1})}=\mbox{\Large{(}}\prod_{n=2}^{N}x_{\alpha}^{1/2}\mbox{\Large{)}}\mbox{   }^{[N-1]}\mbox{He}^{[N]}(\frac{1}{2} ; \frac{3}{2} ; \{x_i\}_{i\in [2,N]})
\eeq
with $\gamma_\alpha$ given by Eq.(27). Eq.(34) can be considered as an integral definition of the multiple hypergeometric function $\mbox{}^{[N-1]}\mbox{He}^{[N]}$ ;  this multiple hypergeometric function is quite complicated, an iterative approximation scheme, called Independent Motions Approximation (IMA), is described in ref.\cite{r7}.\\ 
From the point of view of the physical chemistry, the main purpose of this paper was to compute analytically the Lagrange multipliers for a set of linear holonomic constraints in view of applications to molecular simulations or statistical physics of complex systems. For applications of these analytical results in molecular simulation, the main drawback is that the number of coefficients in the $\tilde{Z}^{-1}$-matrix, that have to be computed and stored, scales as $N^2$ (the matrix is fully filled, see appendix for some examples), $N$ being the number of bonds linearly connected in the molecule. Therefore, for practical implementations, this brute force method of computing Lagrange multipliers is certainly not as efficient as SHAKE or RATTLE algorithms, at least when $N$ is rather large. Another point that one should kept in mind in trying to implement these analytical results in a computer code is that, because of $C_\alpha$, velocities (or momentum) contribute to Lagrange multipliers. This may rise to some technical problems in the Verlet-velocity algorithm and in constant temperature Molecular Dynamics computations, since forces at $t+\delta t$ have to be computed to obtain the velocity at $t+\delta t$ \cite{r1,r2,r4}. Nevertheless, clever uses of these analytical results may certainly be used to improve the efficiency and accuracy of RATTLE algorithms ; for instance, when many bonds verify $\hat{u}_{\alpha-1}\cdot\hat{u}_{\alpha}\simeq 1$.\\
On general grounds, despite the complexity of analytical formulas of $\tilde{Z}^{-1}$ given by Eq.(20), we believe that the analytical results presented in this work are of some interest, since matrix similar to $\tilde{Z}$ appears frequently in many models or theories in one dimension or with nearest neighbors interactions (spin models, harmonic chains, etc.).\\
  
\newpage
\begin{center}
\large{\bf Appendix : Lagrange multipliers for $N=1,2,3,4$ and $5$.}
\end{center}
\renewcommand{\theequation}{A.\arabic{equation}}
\setcounter{equation}{0}
In this appendix, we give explicitly the analytical formulas for linear holonomic constraints that correspond to N=1, 2, 3, 4 and 5 bonds. 
To obtain analytical results for the matrix $\tilde{Z}^{-1}$, we compute firstly $\Omega_n^2$ from the recursive relation in Eq.(17),
\beq
\left\{
\begin{array}{ll}
\displaystyle \Omega_2^2&\displaystyle = 1-\gamma_2^2\\
&\\
\displaystyle \Omega_3^2&\displaystyle =\frac{1-\gamma_2^2-\gamma_3^2}{1-\gamma_2^2}\\
&\\
\displaystyle \Omega_4^2&\displaystyle =\frac{1-\gamma_2^2-\gamma_3^2-\gamma_4^2(1-\gamma_2^2)}{1-\gamma_2^2-\gamma_3^2}\\
&\\
\displaystyle \Omega_5^2&\displaystyle =\frac{1-\gamma_2^2-\gamma_3^2-\gamma_4^2(1-\gamma_2^2)-\gamma_5^2(1-\gamma_2^2-\gamma_3^2)}{1-\gamma_2^2-\gamma_3^2-\gamma_4^2(1-\gamma_2^2)}
\end{array}
\right .
\eeq
The computation of $\tilde{Z}^{-1}$ for $N=1,2,3,4$ and $5$ is done by using Eq.(20).\\[0.2in]
$\bullet\t2\t2\bm{N=1}$\\[0.1in]
\renewcommand{\theequation}{A.1.\arabic{equation}}
\setcounter{equation}{0}
For $N=1$, $\tilde{Z}=\tilde{Z}^{-1}=(1)$ and, following Eqs.(20) and (21), we have
\beq
K_1=\frac{2a}{m}(\bm{F}_1-\bm{F}_0)\cdot\hat{\bm{u}}_1\t2\t2\t2\mbox{and}\t2\t2\t2C_1=\frac{2}{m^2}(\bm{p}_1-\bm{p}_0)^2
\eeq
Then, the Lagrange multiplier is 
\beq
\lambda_1=\frac{1}{2a}\mbox{\large [}(\bm{F}_1-\bm{F}_0)\cdot\hat{\bm{u}}_1+\frac{1}{ma}(\bm{p}_1-\bm{p}_0)^2\mbox{\large ]}
\eeq
the constraint force acting on atom numbered 1 is given by 
\beq
\bm{G}_1=-\lambda_1\bm{\nabla}_{1}\sigma_{1}=-\mbox{\large [}(\bm{F}_1-\bm{F}_0)\cdot\hat{\bm{u}}_1+\frac{1}{ma}(\bm{p}_1-\bm{p}_0)^2\mbox{\large ]}\hat{\bm{u}}_1
\eeq
and the force acting on atom 0 is $\bm{G}_0=-\bm{G}_1$.\\[0.2in]
$\bullet\t2\t2\bm{N=2}$\\[0.1in] 
\renewcommand{\theequation}{A.2.\arabic{equation}}
\setcounter{equation}{0}
For $N=2$, we have
\beq
\tilde{Z}=\left(\begin{array}{cc}1 & -\gamma_2 \\-\gamma_2 & 1\end{array}\right)\t2\t2\t2\mbox{and}\t2\t2\t2\tilde{Z}^{-1}=\frac{1}{1-\gamma_2^2}\left(\begin{array}{cc}1 & \gamma_2 \\ \gamma_2 & 1\end{array}\right)
\eeq
therefore, with $\gamma_2=\hat{\bm{u}}_1\cdot\hat{\bm{u}}_2/2$, Lagrange multipliers are given by 
\beq
\left\{
\begin{array}{lll}
\displaystyle \lambda_1&\displaystyle =\frac{1}{2a(1-\frac{(\hat{\bm{u}}_1\cdot\hat{\bm{u}}_2)^2}{4})}&\displaystyle\mbox{\Large [}(\bm{F}_1-\bm{F}_0)\cdot\hat{\bm{u}}_1+\frac{1}{ma}(\bm{p}_1-\bm{p}_0)^2\\
&&\\
&&\displaystyle +\frac{1}{2}(\hat{\bm{u}}_1\cdot\hat{\bm{u}}_2)\mbox{\large [}(\bm{F}_2-\bm{F}_1)\cdot\hat{\bm{u}}_2+\frac{1}{ma}(\bm{p}_2-\bm{p}_1)^2\mbox{\large ]}\mbox{\Large ]}\\
&&\\
\displaystyle\lambda_2&\displaystyle=\frac{1}{2a(1-\frac{(\hat{\bm{u}}_1\cdot\hat{\bm{u}}_2)^2}{4})}&\displaystyle\mbox{\Large [}\frac{1}{2}(\hat{\bm{u}}_1\cdot\hat{\bm{u}}_2)\mbox{\large [}(\bm{F}_1-\bm{F}_0)\cdot\hat{\bm{u}}_1+\frac{1}{ma}(\bm{p}_1-\bm{p}_0)^2\mbox{\large ]}\\
&&\\
&&\displaystyle +(\bm{F}_2-\bm{F}_1)\cdot\hat{\bm{u}}_2+\frac{1}{ma}(\bm{p}_2-\bm{p}_1)^2\mbox{\Large ]}
\end{array}
\right .
\eeq
The forces acting on atoms are
\beq
\left\{
\begin{array}{ll}
\displaystyle \bm{G}_0 &\displaystyle =-2a \lambda_1\hat{\bm{u}}_1\\
&\\
\displaystyle \bm{G}_1 &\displaystyle =2a \lambda_1\hat{\bm{u}}_1-2a \lambda_2\hat{\bm{u}}_2\\
&\\
\displaystyle \bm{G}_2 &\displaystyle =2a \lambda_2\hat{\bm{u}}_2
\end{array}
\right .
\eeq
$\bullet\t2\t2\bm{N=3}$\\[0.1in] 
\renewcommand{\theequation}{A.3.\arabic{equation}}
\setcounter{equation}{0}
With $N=3$, matrix $\tilde{Z}$ and $\tilde{Z}^{-1}$ are given by
\beq
\tilde{Z}=\left(\begin{array}{ccc}1 & -\gamma_2 & 0 \\-\gamma_2 & 1 & -\gamma_3 \\0 & -\gamma_3 & 1\end{array}\right)
\eeq
and
\beq
 \displaystyle \tilde{Z}^{-1}=\frac{1}{1-\gamma_2^2-\gamma_3^2}\left(\begin{array}{ccc}\displaystyle1-\gamma_3^2 &\displaystyle  \gamma_2 & \displaystyle \gamma_2\gamma_3 \\ &&\\ \gamma_2 & \displaystyle 1  &\displaystyle \gamma_3 \\ &&\\ \gamma_2 \gamma_3 & \gamma_3 & \displaystyle 1-\gamma_2^2 \end{array}\right)
\eeq
\newline
As for $N=1$, following Eqs. (20) and (21), for $1\leq i\leq 3$, $K_i$ and $C_i$ are given by 
\beq
K_i=\frac{2a}{m}(\bm{F}_i-\bm{F}_{i-1})\cdot\hat{\bm{u}}_i\t2\t2\t2\mbox{and}\t2\t2\t2C_i=\frac{2}{m^2}(\bm{p}_i-\bm{p}_{i-1})^2
\eeq
Then using Eqs.(6), (A.3.2) and (A.3.3) we may compute Lagrange multipliers and forces acting on atoms
\beq
\left\{
\begin{array}{ll}
\displaystyle \bm{G}_0 &\displaystyle =-2a \lambda_1\hat{\bm{u}}_1\\
&\\
\displaystyle \bm{G}_1 &\displaystyle =2a \lambda_1\hat{\bm{u}}_1-2a \lambda_2\hat{\bm{u}}_2\\
&\\
\displaystyle \bm{G}_2 &\displaystyle =2a \lambda_2\hat{\bm{u}}_2-2a \lambda_3\hat{\bm{u}}_3\\
&\\
\displaystyle \bm{G}_3 &\displaystyle =2a \lambda_3\hat{\bm{u}}_3
\end{array}
\right .
\eeq
The complicated part of the force being held in Lagrange multipliers.\\[0.2in]
$\bullet\t2\t2\bm{N=4}$\\[0.1in]
\renewcommand{\theequation}{A.4.\arabic{equation}}
\setcounter{equation}{0}
For $N=4$, we have
\beq
\tilde{Z}=\left(\begin{array}{cccc}1 & -\gamma_2 & 0 & 0 \\  -\gamma_2 & 1 &  -\gamma_3 & 0 \\0 &  -\gamma_3 & 1 &  -\gamma_4 \\0 & 0 &  -\gamma_4 & 1\end{array}\right)
\eeq
and the matrix coefficients of $\tilde{Z}^{-1}$ are given by\\[0.3in]
$\displaystyle \tilde{Z}^{-1} = $
\beq
\frac{1}{1-\gamma_2^2-\gamma_3^2-\gamma_4^2(1-\gamma_2^2)}\left(\begin{array}{cccc}\displaystyle 1-\gamma_4^2-\gamma_3^2 & \displaystyle\gamma_2(1-\gamma_4^2) & \displaystyle\gamma_2\gamma_3 & \displaystyle\gamma_2\gamma_3\gamma_4 \\ &&&\\ & 1-\gamma_4^2 & \gamma_3 & \gamma_3\gamma_4 \\  &&&\\ &  & 1-\gamma_2^2 & (1-\gamma_2^2)\gamma_4 \\ &&&\\  &  &  & 1-\gamma_2^2-\gamma_3^2 \end{array}\right)
\eeq
$K_i$ and $C_i$ are given by Eq.(A.3.3), but $1\leq i\leq 4$ and forces by
 \beq
\left\{
\begin{array}{ll}
\displaystyle \bm{G}_0 &\displaystyle =-2a \lambda_1\hat{\bm{u}}_1\\
&\\
\displaystyle \bm{G}_1 &\displaystyle =2a \lambda_1\hat{\bm{u}}_1-2a \lambda_2\hat{\bm{u}}_2\\
&\\
\displaystyle \bm{G}_2 &\displaystyle =2a \lambda_2\hat{\bm{u}}_2-2a \lambda_3\hat{\bm{u}}_3\\
&\\
\displaystyle \bm{G}_3 &\displaystyle =2a \lambda_3\hat{\bm{u}}_3-2a \lambda_4\hat{\bm{u}}_4\\
&\\
\displaystyle \bm{G}_4 &\displaystyle =2a \lambda_4\hat{\bm{u}}_4
\end{array}
\right .
\eeq
where Lagrange multipliers are computed by using Eq.(A.4.2).\\[0.5in]
$\bullet\t2\t2\bm{N=5}$\\[0.1in]
\renewcommand{\theequation}{A.5.\arabic{equation}}
\setcounter{equation}{0}
For $N=5$, the $\tilde{Z}$-matrix is
\beq
\tilde{Z}=\left(\begin{array}{ccccc}1 & -\gamma_2 & 0 & 0 & 0 \\ -\gamma_2 & 1 & -\gamma_3 & 0 & 0 \\0 & -\gamma_3 & 1 & -\gamma_4 & 0 \\0 & 0 & -\gamma_4 & 1 & -\gamma_5 \\ 0 & 0 & 0 & -\gamma_5 & 1 \end{array}\right)
\eeq
and after some algebra using Eq.(20), we find
\beq
\displaystyle \tilde{Z}^{-1} = \frac{1}{1-\gamma_2^2-\gamma_3^2-\gamma_4^2(1-\gamma_2^2)-\gamma_5^2(1-\gamma_2^2-\gamma_3^2)}
\eeq
\newline
$\left(\begin{array}{ccccc}&&&&\\ 1-\gamma_5^2-\gamma_4^2 & \gamma_2(1-\gamma_5^2-\gamma_4^2) & \gamma_2\gamma_3(1-\gamma_5^2) & \gamma_2\gamma_3\gamma_4 & \gamma_2\gamma_3\gamma_4\gamma_5 \\ \t2\t2\t2 -\gamma_3^2(1-\gamma_5^2)&&&&\\ &&&&\\  & 1-\gamma_5^2-\gamma_4^2  & \gamma_3(1-\gamma_5^2) & \gamma_3\gamma_4 & \gamma_3\gamma_4\gamma_5  \\ &&&&\\ &&&&\\  &  & (1-\gamma_2^2)(1-\gamma_5^2) & \gamma_4(1-\gamma_2^2) & \gamma_4\gamma_5(1-\gamma_2^2) \\ &&&&\\ &&&&\\  &  &  & 1-\gamma_2^2-\gamma_3^2& \gamma_5(1-\gamma_2^2-\gamma_3^2) \\ &&&&\\  &&&&\\&   &  &  &1-\gamma_2^2-\gamma_3^2\\ &&&& \t2\t2\t2 -\gamma_4^2(1-\gamma_2^2)\\ &&&& \end{array}\right)$\\[0.2in]
Then, Lagrange multipliers are obtained as previously.\\
The computations done for $N=4$ and 5 show how the complexity of Lagrange multipliers grows with the number of holonomic constraints. On Eqs.(A.4.2) and (A.5.2), the equivalence of both directions of labelling atoms, for linear holonomic constraints, appears explicitly in the structure of the matrix.


\newpage
\vfill
\newpage
\vspace{.5cm}
\begin{thebibliography}{99}

\bibitem{r1} M.P. Allen and D.J. Tildesley, \it{Computer\rst1 Simulation\rst1 of\rst1 Liquids} (Clarendon Press, Oxford,1987)   
\bibitem{r2} H.C. Andersen,  $J.$ $Comput.$ $Phys.$ $\bm{52}$, 24 (1983) 
\bibitem{r3} S.W. de Leeuw, J.W. Perram and H.G. Petersen, $J.$ $Stat.$ $Phys.$ $\bm{61}$, 1203 (1990) 
\bibitem{r4} C. Pierleoni and J.-P. Ryckaert, $Mol.$ $Phys.$, $\bm{75}$, 731 (1992) 
\bibitem{r5} M. Fixman,  $Proc.$ $Nat.$ $Acad.$ $Sci.$ $USA$ $\bm{71}$, 3050 (1974)
\bibitem{r6} S. Melchionna, $Phys.$ $Rev.$ E  $\bm{61}$, 6165 (2000)
\bibitem{r6a} G.R. Kneller, $J.$ $Chem.$ $Phys.$, $\bm{125}$, 114107 (2006)
\bibitem{r7} M. Mazars, $J.$ $Phys.$ $A:$ $Math.$ $Gen.$ $\bm{31}$, 1949 (1998) 
\bibitem{r8} M.E. Tuckerman, C.J. Mundy and G.J. Martyna $Europhys.$ $Lett.$ $\bm{45}$, 149 (1999)  
\bibitem{r9} M. Fixman and J. Kovac, $J.$ $Chem.$ $Phys.$, $\bm{61}$, 4939 (1974) ; $ibid$, 4950 (1974)
\bibitem{r10} M. Mazars, $Phys.$ $Rev.$ E  $\bm{53}$, 6297 (1996) 

\end{thebibliography}
\end{document}